\documentclass[prl,floatfix,reprint,twocolumn]{revtex4}
\usepackage{graphicx,amssymb,amsmath}
\usepackage{url}
\usepackage{hyperref}  
\usepackage{courier}

\begin{document}

\title{A global take on congestion in urban areas}

\author{Marc Barthelemy}
\email{marc.barthelemy@cea.fr}
\affiliation{Institut de Physique Th\'{e}orique, CEA, CNRS-URA 2306, F-91191, Gif-sur-Yvette, France}
\affiliation{Centre d'Analyse et de Math\'ematique Sociales, EHESS-CNRS (UMR 8557),190-198 avenue de France, FR-75013 Paris, France}

\begin{abstract}
We analyze the congestion data collected by a GPS device company (TomTom) for almost 300 urban areas in the world. Using simple scaling arguments and data fitting we show that congestion during peak hours in large cities grows essentially as the square root of the population density. This result, at odds with previous publications showing that gasoline consumption decreases with density, confirms that density is indeed an important determinant of congestion, but also that we need urgently a better theoretical understanding of this phenomena. This incomplete view at the urban level leads thus to the idea that thinking about density by itself could be very misleading in congestion studies, and that it is probably more useful to focus on the spatial redistribution of activities and residences. 
\end{abstract}

\maketitle

The increasing likelihood that an ever larger number of urban inhabitants can afford a private car continues to structure the spatial organization of our cities \cite{Anas:1998} with dramatic effects on their efficiency and development. Even when urban infrastructures are drastically remodelled in favour of the automobile, congestion keeps growing and has become one of the most important challenge for politicians and planners. In addition, it is an ever more important cause of serious health problems \cite{DeWeerdt:2016}, and congestion leads to significant loss of time through traffic jams which has a negative impact on the economical growth of cities.

With new sources of data, we are hard at work in reaching out for a quantitative understanding of cities \cite{Batty:2013} and this science helps to make these statements about congestion more precise. In particular, the data provided (now on a yearly basis) by one of the major GPS navigation device companies \cite{tomtom:2016} allows us to produce a certain number of results about congestion in world cities that are worth reflecting upon. Of particular interest is the estimate of extra travel time per day $\delta$ due to congestion. It is obtained by computing the increase of the average travel times during peak hours compared to a free flow situation. For London the extra travel time per day is $39$ minutes for a 1 hour trip, and this can be compared to Mexico ($57$'), Los Angeles ($43$'), Beijing ($42$'), Paris ($38$') and Johannesbourg ($35$') which are immediate examples computed from such data. In other words, if a trip in free flow (without congestion) has a duration $\tau_0$, with congestion (during peak hours) it will take a time equal to $(1+\delta)\tau_0$ (where $\delta$ and $\tau_0$ are measured in units of one hour).

This information allows us to discuss regional peculiarities and to monitor the yearly increases in congestion \cite{Andersen:2016,Cox:2016,Pisarski:2016}. In principle, it also allows us to compare different cities with one other, but this has however to be done with care, and it could be misleading to compare the extra travel time per day directly. Indeed, with different sizes of city, a one hour trip could be close to the average duration of trips in one size of city, whereas in a smaller city, it could be above the average: thus the average commuting time clearly depends on the size of the city. For example, in the US \cite{Sivak:2015}, the average commuting time is of order $26$ minutes and varies from $40$ minutes in New York City to $23$ minutes in Indianapolis while we also note that generally speaking commuting times are not stable over time and depend on urban spatial structure \cite{Levinson:2005} (see also \cite{Anas:2015} for a detailed discussion of the Chicago case). A value of $\delta=10$ minutes, for example, has therefore a very different meaning in a city where most trips have a duration of $10$ minutes in comparison to a city where the average is one hour. In the former case, congestion represents $50\%$ of the trip of total duration $20$ minutes, while in the latter it represents only about $14\%$ of the total trip duration. Unfortunately, data about the average commuting time is usually not available for cities in different parts of the world, and if we want to compare the effect of congestion in different cities, we have to identify a typical trip duration in each city.

There are many length and time scales in a city such as those associated with different transportation modes or the cost incurred -- the  financial aspects of transport \cite{Louf:2014}, but an important determinant is a city's area $A$ which gives the order of magnitude $\sim\sqrt{A}$ for the length of a typical trip in the city. If we assume that the average free flow speed $\overline{v}$ is constant, the typical free flow trip duration is given by $\tau_0=\sqrt{A}/\overline{v}$ and the total delay (for the whole population $P$ of the city) due to congestion (during peak hours) is given by
\begin{equation}
\Omega=P\delta\tau_0
\end{equation}
This quantity $\Omega$ represents the total delay experienced by the population of the urban area and constitutes an upper bound to the time lost in congestion as we consider here that the whole population is travelling by car and experiences the same average delay.
\begin{figure}[!h]
	\centering
	\includegraphics[width=0.49\textwidth]{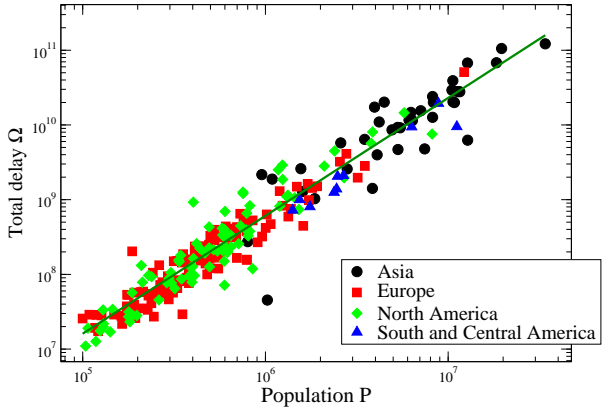}
	\caption{ Total delay (up to a constant factor $\overline{v}$) due to traffic jams in cities versus their population. The different symbols correspond to cities in different regions of the world. The straight line represents a power law fit $\Omega\sim P^\beta$ with exponent $\beta\approx≈1.58$ ($r^2=0.96$). The data for the extra travel time is from Tomtom \cite{tomtom:2016}; the data for the area and population are measured for urban areas and are from the United Nations \cite{UN}.}
\label{fig:1}
\end{figure}

We show in Figure 1 this total delay (up to a factor $\overline{v}$) versus the population and this displays a clear scaling \cite{Pumain:2004,Bettencourt:2007} of the form $\Omega=\omega_0P^\beta$ with an exponent about $1.58$ and a prefactor $\omega_0=0.21$hour. We note that a direct power law fit on the extra travel time per day gives $\delta\sim P^{0.17}$.  These results imply that the total delay increases quickly with the population, and that the extra travel time per day scales dominantly as
\begin{equation}
\delta\sim \frac{\overline{v}\omega_0P^{\beta-1}}{\sqrt{A}}\approx \overline{v}\omega_0\sqrt{\rho}
\end{equation}
where $\rho=P/A$ is the average population density of the urban area (Note that all these results are consistent with each other, given that the area scales with population as $A\sim P^{0.81}$ for cities in this dataset). We also see that there is a small logarithmic-like correction to this behavior Eq. (2) of order $P^{0.08}$ (which is of the order of unity), and the determinant factor of the extra travel time is the density of the urban area. It is interesting to note that Eq. (2) can be seen as the ratio of the average free flow velocity and another velocity given by the displacement over a distance of order $1/\sqrt{\rho}$ for a `universal' duration $\omega_0\approx 13$ minutes. The fact that the main determinant for congestion seems to be the density is consistent with previous suggestions \cite{Cox:2014}, and this `slow' square root behavior is rather good news. Also, this result from Eq. (2) shows that the extra travel time per day cannot be used for a direct comparison between cities but their difference in terms of density should be taken into account.

On the financial side which involves many costs  such as transport, health- and business-related which are associated to congestion,  accounting for all of them can be difficult. We can estimate a congestion-related cost in a simple way by converting the lost time in traffic jams during peak hours into a financial cost using the average hourly income $y$ in the country which the city belongs to. We can then define the ‘financial loss’ $\eta$ due to traffic jams as a percentage of the GDP per capita $g$ of the country by
\begin{equation}
\eta=\frac{\delta\tau_0 y}{g}
\end{equation}

We show this quantity versus population in Figure 2 and we observe a significant increase of this financial loss with population from less than $1\%$ to almost $10\%$ of the GDP. 
\begin{figure}[!h]
	\centering
	\includegraphics[width=0.49\textwidth]{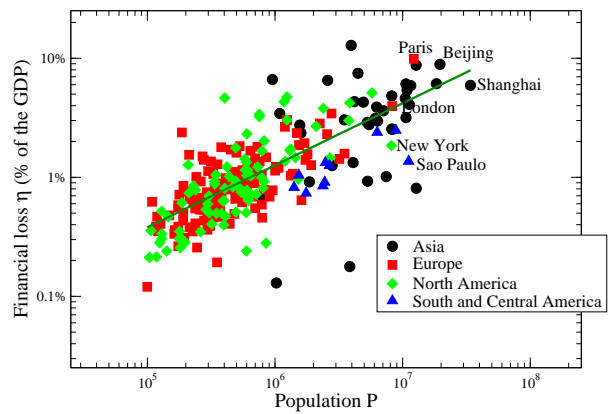}
	\caption{Financial loss due to congestion in percent of the gdp per capita (computed for $\overline{v}=50$km/h). 
The straight line is a power law fit with exponent $0.52$ ($r^2=0.73$).}
\label{fig:2}
\end{figure}
A power law fit gives a behavior close to $\eta\sim\sqrt{P}$ indicating that even if congestion increases slowly with population, it generates a non-negligible cost. Asian cities have a larger loss (on average $4\%$) followed by the other regions which have roughly the same level (of order $1-2\%$).  

There is however a large dispersion around this average behaviour which shows that the population is not the only determinant of this cost. More precisely, for a given value of the population, we observe large differences which probably reflect the efficiency of road infrastructures. We can compute the deviation from the average behavior defined by the power law fit shown in Figure 2 and we observe that in the group of megacities, Paris conjugates both a large income and a high level of congestion, while in Sao Paulo average income is low, leading to a relatively small loss. In Beijing, the average income does not compensate the high congestion level, while in London which has a population and an average income of the same order as Paris, we observe an average loss.

What can we conclude from these different observations ? First of all, for the world cities studied in this Tomtom (2016) dataset, congestion obeys a scaling law which seems quite independent from the cultural and historical differences between these cities. Congestion grows -- relatively slowly -- with the average density and displays an apparent square root behavior. This is the sign that a fundamental mechanism is at play here and begs for some theoretical modeling. In addition, this increase is at odds with the older observation that gasoline consumption decreases with the average density \cite{Newman:1989}, and suggests that compact development cannot be systematically associated with a decrease in congestion \cite{Echenique:2012}. We thus have contradictory results with respect to density which has an unclear role and this shows that our theoretical understanding of congestion at the urban level, at least,  is incomplete. We need more measures and also new theoretical insights in order to understand the impact of density on congestion, pollution and gasoline consumption. When facing this puzzle, it seems difficult to provide to urban planners and policy makers with good scientific advice which is grounded in observation and theory.  However,  we understand here that there is at least two factors which play a major role. The first one is obviously the share of individuals travelling to work by car: a decreasing share would decrease $\Omega$ (but is not likely to change its scaling). Public transportation is a good alternative, especially if it is not too sensitive to congestion. Also, commuting distance is another key factor that governs the duration of the trip: decreasing $\tau_0$ will also decrease $\Omega$. In some countries (e.g. US, UK, Denmark), this distance is broadly distributed \cite{Carra:2016} which suggests that we are far from an optimal spatial organization with many mixed land-use centers (such as `urban villages') scattered throughout the city.

Thinking about density by itself could thus be very misleading in congestion studies, and it is probably more useful to focus on the spatial redistribution of activities and residences. This could give some clues on how we might reduce the fraction of car users and commuting distances, and how we might develop and encourage healthier transportation modes.

\end{document}